\documentclass[]{emulateapj}

\def\bibfiles{biblio1}
\def\aareferences{\bibliographystyle{aa}
                  \bibliography{\bibfiles}}

\def\rmit#1{{\it #1}}              
\def\specchar#1{{\sc #1}}
\def\FeI{\mbox{Fe\,\specchar{i}}}

\def\SiI{\mbox{Si\,\specchar{i}}}
\def\HeI{\mbox{He\,\specchar{i}}}
\def\CaIIH{\mbox{Ca\,\specchar{ii}\,\,H}}
\def\CaII{\mbox{Ca\,\specchar{ii}}}

\def\ie{\rmit{i.e.}}
\def\eg{\rmit{e.g.}}

\def\apj{ApJ}

\usepackage{natbib,graphicx}

\usepackage{color}

\begin{document}

\title{Multi-layer study of wave propagation in sunspots}



\author{T. Felipe\altaffilmark{1,2}, E. Khomenko\altaffilmark{1,2,3}, M. Collados\altaffilmark{1,2} and C. Beck\altaffilmark{1,2}}

\altaffiltext{1}{Instituto de Astrof\'{\i}sica de Canarias, 38205,
C/ V\'{\i}a L{\'a}ctea, s/n, La Laguna, Tenerife, Spain}
\altaffiltext{2}{Departamento de Astrof\'{\i}sica, Universidad de La Laguna, 38205, La Laguna, Tenerife, Spain}
\altaffiltext{3}{Main Astronomical Observatory, NAS, 03680, Kyiv,
Ukraine}

\begin{abstract}
We analyze the propagation of waves in sunspots from the
photosphere to the chromosphere using time series of co-spatial
\CaIIH\ intensity spectra (including its line blends) and
polarimetric spectra of \SiI\ $\lambda$ 10827 and the \HeI\
$\lambda$ 10830 multiplet. From the Doppler shifts of these lines
we retrieve the variation of the velocity along the line-of-sight
at several heights. Phase spectra are used to obtain the relation
between the oscillatory signals. Our analysis reveals standing
waves at frequencies lower than 4 mHz and a continuous
propagation of waves at higher frequencies, which steepen into
shocks in the chromosphere when approaching the formation height of the \CaIIH\ core.
The observed non-linearities are weaker in \CaIIH\ than in \HeI\
lines. Our analysis suggests that the \CaIIH\ core forms at a
lower height than the \HeI\ $\lambda$ 10830 line: a time delay of
about 20 s is measured between the Doppler signal detected at both
wavelengths. We fit a model of linear slow magnetoacoustic wave
propagation in a stratified atmosphere with radiative losses
according to Newton's cooling law to the phase spectra and derive
the difference in the formation height of the spectral lines. We
show that the linear model describes well the wave propagation up
to the formation height of \CaIIH, where non-linearities start to
become very important.
\end{abstract}

\keywords{Sun: magnetic fields; Sun: oscillations}
\maketitle


\section{Introduction}

Since the first detection of waves in sunspots
\citep{Beckers+Tallant1969}, many studies have been carried out to understand the physics of these waves, from the
observational as well as theoretical point of view. Waves in
sunspots are different from those observed in quiet Sun due to the
presence of the magnetic field. They show a variety of behavior
depending on the height and the region of the sunspot where they
are observed. Usually, these waves are classified as photospheric
umbral oscillations, chromospheric umbral oscillations, and
running penumbral waves \citep{Lites1992}. However, all these kind
of waves seem to be different manifestations of the same global
propagation of magneto-acoustic waves \citep{Rouppe+etal2003}. The
reasons to study waves extend beyond the derivation of their
properties, because waves can also be used as independent
diagnostic of sunspot structure \citep{Duvall+etal1996}.

At the photosphere in the umbra, the power spectra of the
oscillations are quite similar to the corresponding ones of the
quiet Sun, with a broad distribution of frequencies and a clear peak at about
3 mHz \citep{Thomas+etal1982}. In sunspots, these oscillations in the 5 min
band have reduced amplitudes compared to the surrounding quiet photosphere
\citep{Abdelatif+etal1986}. The magnetic field also modifies the $p-$modes and
produces new modes of fluctuations not present in the quiet Sun
\citep{Cally+Bogdan1997, Khomenko+etal2009}. Generally, most of the
  photospheric umbra is covered by coherent oscillations
  \citep{Kobanov1990}. Waves are usually studied by measuring the
fluctuations of intensity and line-of-sight velocity from Doppler
shift, but they are also supposed to affect magnetic field. From
full Stokes inversions, \citet{Lites+etal1998} found an upper
limit of 4 G for the amplitude of 5 minute oscillations in
magnetic field strength, while \citet{BellotRubio+etal2000}
detected variations around 7-11 G. In a later work,
\citet{Khomenko+etal2003} interpreted magnetic field oscillations
as being due to fast and slow magneto-hydrodynamic (MHD) wave
modes, producing both intrinsic variations of the magnetic field
and those due to the shift of the line forming region.

The chromosphere of sunspots is dominated by 3 min oscillations.
The power spectra usually show sharp peaks around 5-6 mHz, and the
power drops gradually to noise values in the interval from 8 mHz
to 10-15 mHz \citep{Lites1984}. Several spectral lines, formed at
different heights from the photosphere to the chromosphere, can be
used to sample the wave propagation properties. Simultaneous
time-series of these spectral lines are a powerful tool for
studying sunspot waves. From phase spectra between the umbral
oscillations observed in the photospheric line \FeI\ $\lambda$
5233 and in H$\alpha$, \citet{Giovanelli+etal1978} found that the
phase delay indicates upward wave propagation. \citet{Lites1984}
inferred that slow mode waves propagate upward into the
chromosphere in the frequency band around 6.5 mHz, based on the
phase differences between the oscillations of \CaII\ $\lambda$
8498, \CaII\ $\lambda$ 8542 and \FeI\ $\lambda$ 5434. From the
study of the Doppler shifts observed in the intensity profiles of
the \HeI\ $\lambda$ 10830 multiplet, \citet{Lites1986} presented
evidence of shock formation at the chromosphere.
\citet{Centeno+etal2006} reproduced the phase spectra between
chromospheric and photospheric velocity oscillations with a model
of linear vertical propagation of slow magnetoacoustic waves in a
stratified magnetized atmosphere that accounts for radiative
loses, finding a good agreement between the theoretically computed
time delay, and the one obtained from the cross-correlation of
photospheric \SiI\ $\lambda$ 10827 and chromospheric \HeI\
$\lambda$ 10830 velocity maps, filtered around the 6 mHz band.
They showed that the chromospheric 6 mHz signal is a result of
linear wave propagation of the photospheric perturbations in the 6
mHz range, rather than the consequence of the nonlinear
interaction of photospheric modes as proposed by
\citet{Gurman+Leibacher1984}.


The works cited above were limited to the study of oscillations
at only two heights (one photospheric and one chromospheric),
separated by several hundreds of kilometers. It is thus
interesting to perform a more detailed sampling of the sunspot
atmosphere, using more spectral lines formed at several
intermediate heights between these two regions. On the one hand,
observationally detected spatial wave patterns in sunspots are
rather complex \citep{Bogdan+Judge2006}. On the other hand, recent
numerical simulations of waves in sunspots also suggest a complex
picture of the fast and slow magneto-acoustic waves propagating
simultaneously in the same layers but in different directions and with different phase speeds \citep{Khomenko+Collados2009}. This requires a more refined
multi-layer study of sunspot waves. Studies of this kind often
represent an observational challenge since several spectral lines
have to be detected simultaneously not only in intensity but also
in polarized light. In our paper, we report on such multi-line
spectropolarimetric observations. Our aim is to cover the gap
between the photospheric and chromospheric signals and analyze
sunspot oscillations at the formation heights of several spectral
lines formed between \SiI\ and \HeI. For that we use simultaneous
observations obtained with two instruments, the POlarimetric LIttrow Spectrograph
\citep[POLIS,][]{Beck+etal2005b} and the Tenerife Infrared
Polarimeter II \citep[TIP-II,][]{Collados+etal2007},
attached to the German Vacuum Tower telescope at the Observatorio
del Teide at Tenerife. Apart from information about waves, our
multi-layer study has also allowed us to estimate the formation heights of the spectral lines in the sunspot atmosphere, including
\CaIIH\ line, several \FeI\ blends in the wing of this line and
the infrared lines of \SiI\ at $\lambda$ 10827 \AA\ and \HeI\ at
$\lambda$ 10830 \AA.

The structure of the paper is the following. In
Sect.~\ref{sect:observation}, the observations and data reduction
are explained. Section \ref{sect_ana} describes the analysis of
the velocity oscillations at several heights.  The
results are discussed in  Sect.~\ref{sect_disc}, which also presents
our conclusions.

\section{Observation and data reduction}
\label{sect:observation} The observations analyzed in this work
were obtained on 2007 August 28 with two different instruments,
POLIS and TIP-II, attached to
the German Vacuum Tower Telescope (VTT) at the Observatorio del
Teide. Simultaneous and co-spatial scans of a sunspot located near
the center of the Sun ($\mu=0.96$) were taken with both
instruments. The slit was placed over the center of the sunspot.
The observations were obtained with real-time seeing correction by
the Kiepenheuer-Institute adaptive optics system
\citep{vonderluehe+etal2003}.

The spectra of the blue channel of POLIS include \CaIIH\ $\lambda$
3968 \AA\ intensity profiles and some photospheric line blends in
the wings of \CaIIH, covering a spectral range from 3964.9 \AA\ to
3971.3 \AA\ with a spectral sampling of 20 m\AA\ pixel$^{-1}$ and
a spatial sampling of $0''29$ per pixel. The Ca spectra were
reduced for the flatfield \citep{Beck+etal2005a, Beck+etal2005b},
and were also corrected for the transmission curve of the
order-selecting interference filter in front of the camera. For
the wavelength calibration, the line-core positions of the iron
lines at 3965.45, 3966.07, 3966.63, 3967.42 and 3969.26 \AA\ in an
average quiet Sun region were determined by a second order
polynomial fit. We then determined the wavelength scale that
matched best all the position values.
\begin{figure}
\centering
\includegraphics[width=9cm]{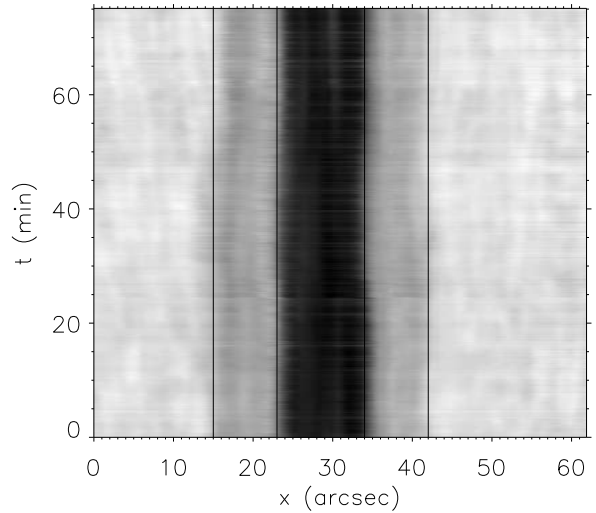}
\caption{Intensity of IR continuum. Vertical lines mark the
position of quiet Sun-penumbra and penumbra-umbra boundaries. Time
increases from bottom to top.} \label{fig:intensity}
\end{figure}

TIP-II yielded the four Stokes parameters $IQUV$ in a spectral
range from 10822.7 \AA\ to 10833.7 \AA\, with a spectral sampling
of 11 m\AA\ pixel$^{-1}$ and a spatial sampling of $0''18$ per
pixel. This spectral region contains information about two
different heights of the solar atmosphere due to the presence of
two spectral lines. The \SiI\ line at 10827.09 \AA\ is formed at
photospheric heights, whereas the \HeI\ $\lambda$ 10830 triplet,
which includes a weak blue component at 10829.09 \AA\ and two
blended red components at 10830.25 and 10830.34 \AA, forms in the
chromosphere \citep{Centeno+etal2008}. In this case, the
wavelength calibration was done using the \SiI\ and \HeI\ lines as
references.
\begin{figure*}
\centering
\includegraphics[width=18cm]{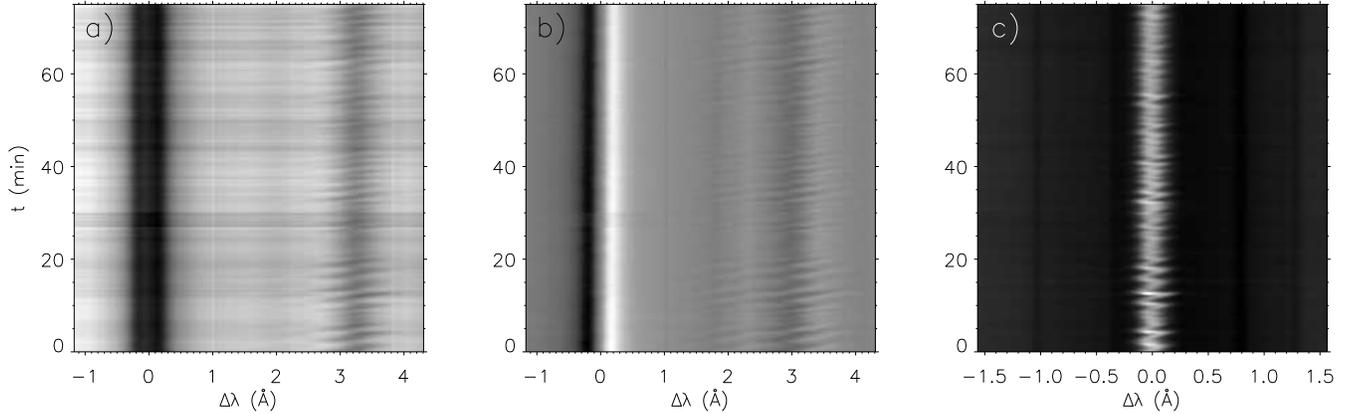}
\caption{Temporal evolution at one position in the umbral region
of series \#2.  \emph{Left}: \SiI\ and \HeI\ intensity;
\emph{center}: \SiI\ and \HeI\ Stokes $V$; \emph{right}: \CaIIH\
intensity. The horizontal axis represents wavelength, with the
origin at the position of the \SiI\ $\lambda$ 10827 \AA\ rest
wavelength in (a) and (b) and at the position of the \CaIIH\
$\lambda$ 3968 \AA\ rest wavelength in (c). The vertical axis
represents time, increasing from bottom to top.}
\label{fig:spectra}
\end{figure*}

The full data set consists of three temporal series; each of them
lasting about one hour. Three scan steps with $0''5$ step width
were taken for the two first series, while in the last series only
two spatial positions were used. The cadence was different for all
series. Table \ref{tb:observation} shows the time step between two
spectra taken at the same spatial position, the time when the
observation were obtained and the number of repetitions of each
series.

\begin{table}[htbp]
  \begin{center}
  \centering \caption{Summary of observations:}
  \label{tb:observation}
  \smallskip
  \begin{tabular}{ccccc}
\hline
         & t$_{start}$ (UT)  & t$_{end}$ (UT)  & $\Delta$t (s) & N spectra\\
\hline
Series 1  & 07:55:02          & 08:54:48        &  21            & 170\\
Series 2  & 09:01:02          & 10:15:25        &  18            & 250 \\
Series 3  & 10:43:34          & 11:58:43        &  7.5           & 600 \\
\hline
 \end{tabular}
  \end{center}
 \end{table}

\begin{figure}
\centering
\includegraphics[width=9cm]{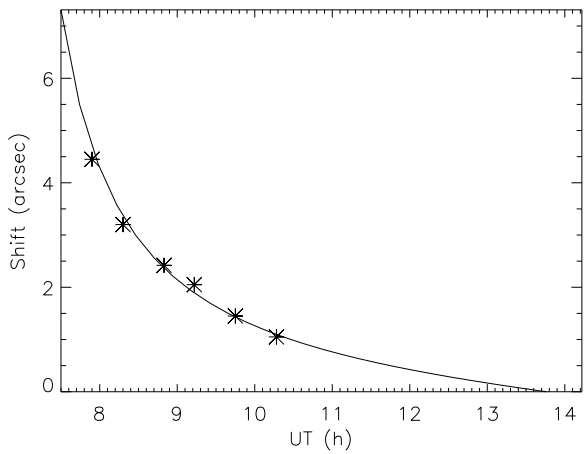}
\caption{Temporal variation of the theoretical spatial
displacement perpendicular to the slit due to differential
refraction \emph{(Solid line)}. \emph{Asterisks} mark the values
measured at the telescope.} \label{fig:refraction}
\end{figure}

Due to the differential refraction in the earth atmosphere
\citep[e.g.,][]{Filippenko1982}, the spectra of POLIS and TIP-II
are not fully co-spatial. The spatial displacement of the two
wavelengths (3968 \AA, 10830 \AA) perpendicular to the slit
depends on the date and time of the observations, the slit
orientation, and the location of the first coelostat mirror
\citep[see Appendix A of][]{Beck+etal2008}. On the first day of
the observation campaign, we took a set of large-area scans at
different times for an accurate determination of the displacement.
The solid line in Fig.~\ref{fig:refraction} shows the
theoretically predicted spatial displacement perpendicular to the
slit due to differential refraction, while the asterisks are the
measured displacements; the match between both is remarkable. To
guarantee an overlap between the observations in the two
wavelengths, we thus positioned the scan mirror inside of POLIS at
the beginning of each observation such that it compensated the
spatial displacement for about the middle of the observation.
Moreover, small repeated scans of 2-3 slit spatial positions,
separated by $0''5$, were taken in order to sample a wider region
of the Sun and prevent possible errors between the theoretical
differential refraction and the actual one. 

Figure \ref{fig:intensity} shows an intensity map from a
wavelength in the infrared continuum. The regions of quiet Sun,
penumbra, and umbra of the sunspot are well defined, the vertical
lines indicate the boundaries between these areas. The temporal
evolution of the Stokes $I$ and Stokes $V$ spectra for the TIP
data and the intensity around the \CaIIH\ core in the POLIS data are
plotted in Fig. \ref{fig:spectra} at a fixed position inside the
umbra. The Stokes $I$ and $V$ profiles from TIP contain the \SiI\
line at $\Delta\lambda=$0 \AA\ (the rest wavelength of the silicon
line was determined from the quiet Sun region and was set as the
origin) and the \HeI\ line at $\Delta\lambda=$3.25 \AA. The helium
line profile shows periodic shifts with large displacements to the
blue and the red. The core of the \CaIIH\ line shows a strong
emission peak inside the umbra. The Doppler shift of this emission
peak develop a saw-tooth pattern, similar to the \HeI\ line
\citep[see also][]{Rouppe+etal2003}.

\section{Data analysis and results}
\label{sect_ana} In this paper, we focus on the line-of-sight
(LOS) velocities. For all the spectral lines besides \CaIIH,
Doppler velocities were inferred by measuring the position of the
intensity minimum. The wavelengths close to the core of the line
were fitted with a second order polynomial. The location of the
minimum of the parabola was taken as the line-core position. This
procedure was performed to obtain the Doppler shifts of \SiI,
\HeI, and the \FeI\ lines at 3965.45, 3966.07, 3966.63, 3967.42
and   3969.26 \AA. In the case of the TIP data, the Doppler shift
of the Stokes $V$ zero crossing, where the polarization signal
intersects the zero level, was derived as well. The Doppler shifts
from the intensity and Stokes $V$ profiles are very similar due to
the large magnetic filling factor in the sunspot umbra.

\begin{figure}
\centering
\includegraphics[width=9cm]{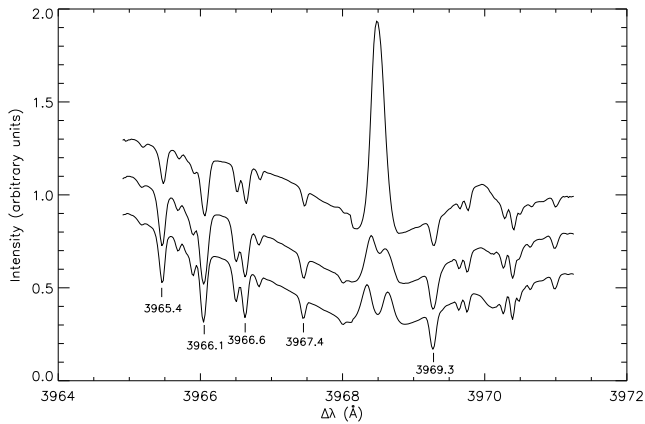}
\caption{Averaged intensity profile of the \CaIIH\ line at
different locations. From bottom to top: quiet Sun, penumbra, and
umbra. The intensity scale is different for all of them. The
profiles have been displaced from each other. The \FeI\ lines analyzed in this work are marked.} \label{fig:profile_Ca}
\end{figure}

The behavior of the \CaIIH\ core is different. It exhibits a
prominent peak at the center of the line in highly magnetized
regions (top line of Fig. \ref{fig:profile_Ca}), while in a
field-free atmosphere the center of the line has a minimum between
two lobes at both sides with their corresponding maxima (bottom
line of Fig. \ref{fig:profile_Ca}) \citep[see also,
e.g.,][]{Liu+Smith1972}. In the umbra, the Doppler shift was
retrieved from the spectral displacement of the maximum of the
core emission. In the low magnetized, region it was obtained from
the shift of the central minimum. All in all, we have obtained
maps of the LOS velocity at each spatial point covered by the slit
at the formation heights of the eight spectral lines, except for
\HeI, whose line depth in the non-magnetized regions is too small
to determine its location.

We have selected the scan steps which give the best alignment of the data from both instruments to obtain a better correction of the differential refraction. The use of different steps introduce a systematic delay between the oscillatory signals from POLIS and TIP-II, since the same spatial location is observed with a time lag. However, in the case of the data from series 3, the theoretical correction of the differential refraction worked perfectly and the optimal alignment was found using the simultaneous scan steps. This data set has been chosen to calculate the phase difference spectra from Section \ref{sect:phase_spectras}. An additional finer alignment along the
slit between TIP and POLIS was done by a cross-correlation of the velocity maps of two spectral lines formed at
similar heights, using the line pair \CaIIH\ and \HeI\ for the
chromosphere, and \SiI\ and \FeI\ $\lambda$ 3969.26 \AA\ for the
photosphere. Finally, we resampled all velocity maps with the
sampling of the POLIS \CaIIH\ data, \ie, $0''29$ per pixel.

\subsection{LOS velocity}

\begin{figure*}
\centering
\includegraphics[width=18cm]{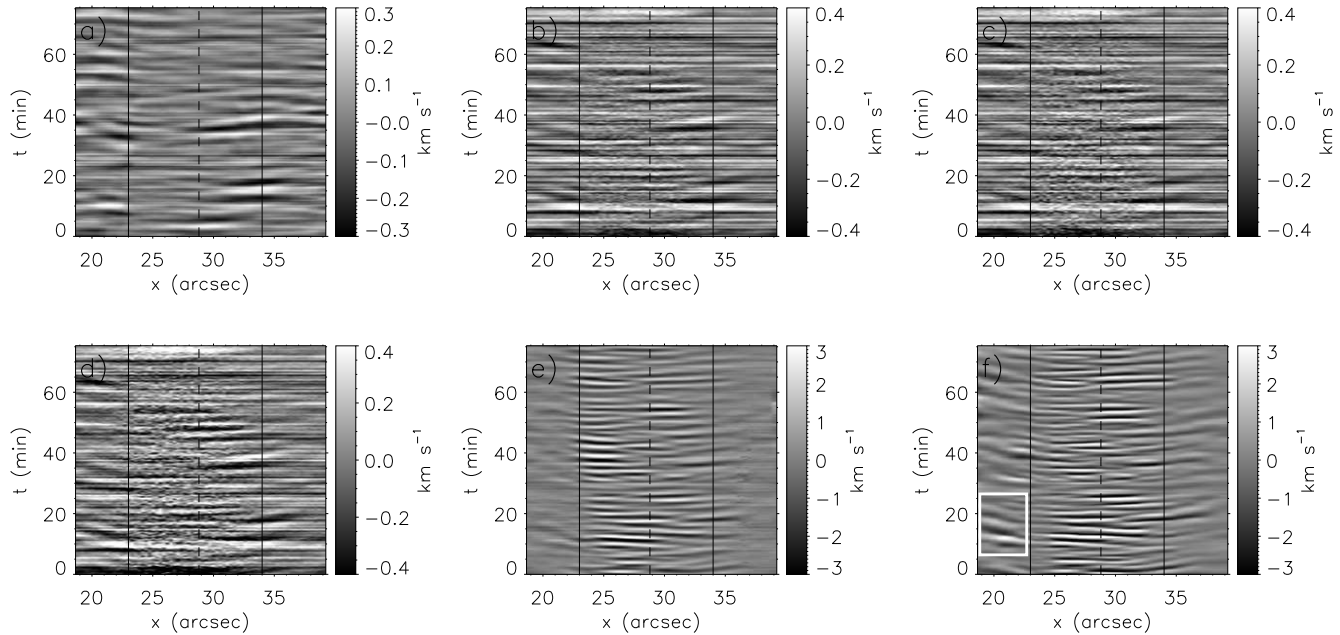}
\caption{Velocity maps in the sunspot during series \#2 for the
different spectral lines. The horizontal axis represents the
position along the slit and the vertical axis represents time.
Black color indicates negative velocities (upflow), white color
positive velocities (downflow). Top row, left to right: \SiI\
$\lambda$ 10827 (a), \FeI\ $\lambda$ 3966.1 (b), \FeI\ $\lambda$
3966.6 (c). Bottom row, left to right:  \FeI\ $\lambda$ 3969.3
(d), \CaIIH\ 3968.5 (e), \HeI\ $\lambda$ 10830 (f). Vertical solid
lines represents the limits of the umbra. The vertical dashed line
denotes the location of the spectra shown in Fig.
\ref{fig:velocity_x}. The white rectangle in (f) marks an area of
running penumbral waves.} \label{fig:velocity_maps}
\end{figure*}

Figure \ref{fig:velocity_maps} shows the temporal evolution of the
LOS velocities in the sunspot region obtained from the Doppler
shift of several spectral lines, sorted by formation height.
Negative velocities (appearing as black shaded regions) indicate
upflows, where matter is approaching the observer, while white
regions are downflows. Figures \ref{fig:velocity_maps}(a-d) reveal
a similar pattern since all these lines are formed at photospheric
heights. The velocity maps of the \FeI\ line blends of \CaIIH\
have lower quality than the \SiI\ maps, especially inside the
umbra, due to the coarser spectral sampling and the small line
depth of the lines.

The velocity field at chromospheric heights is given in Fig.
\ref{fig:velocity_maps}e (\CaIIH) and \ref{fig:velocity_maps}f
(\HeI). Both figures show a similar wave pattern and similar
amplitudes, but, as a more detailed analysis will reveal, the
velocity oscillations of the \HeI\ line are slightly retarded
relative to those in Ca. The wave pattern measured at the
chromosphere in the umbral region has a smaller spatial coherence
than that observed in the photosphere. It also differs in its
larger peak-to-peak variations of about 8 km s$^{-1}$ and its
period of about three minutes.

In the region of the penumbra, which can be seen in Fig.
\ref{fig:velocity_maps}f at a position between $19''$ and $23''$
and between $34''$ and $39''$, there is also a characteristic
pattern of alternating positive and negative velocities, but with
lower amplitudes than the waves in the umbra and with longer
periods, corresponding to running penumbral waves \citep[see, e.g.,][]{Giovanelli1972, Zirin+Stein1972,Bloomfield+etal2007b}. They start at
the inner penumbra, and their wave front is delayed in the regions
of the outer penumbra, so their propagation through the penumbra
appears as a slope in the diagram. An example of such a slope can
be seen in an area between $19''$ and $23''$ during the first 22
minutes (white rectangle). This slope is smaller near the outer
penumbra, indicating a decrease of the propagation speed as the
disturbance travels from the inner to the outer penumbral
boundary. The estimation of the velocity for the \CaIIH\ line core
(Fig. \ref{fig:velocity_maps}e) in the penumbral region is poorer
due to the change of the shape of the line that does not have a
prominent emission peak any more (see middle curve in Fig.
\ref{fig:profile_Ca}).

\begin{figure}
\centering
\includegraphics[width=8.cm]{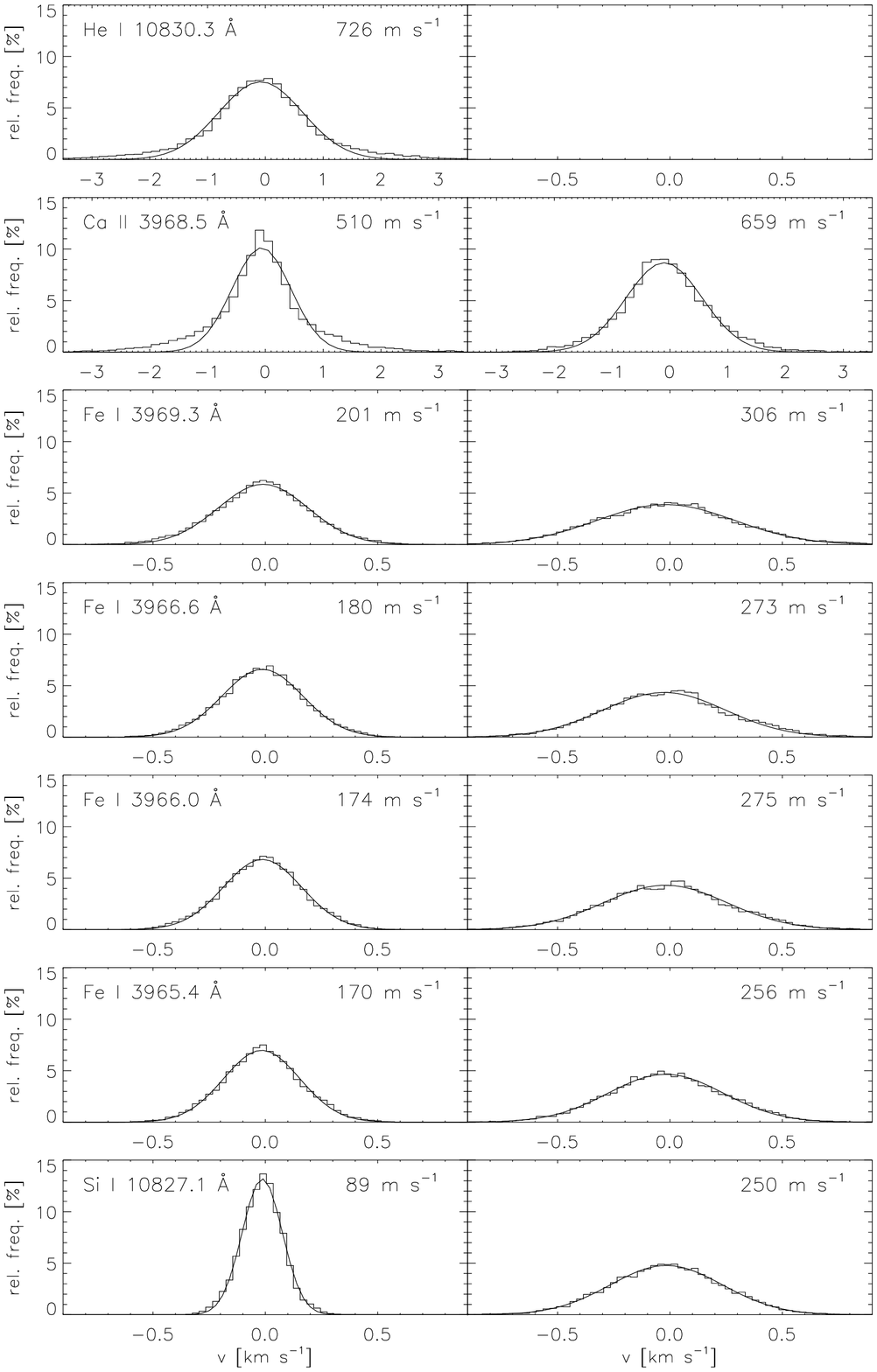}
\caption{Histograms of LOS velocity measured with several spectral
lines. The left column corresponds to the umbra, the right column
to the quiet Sun. From bottom to top: \SiI, \FeI\ $\lambda$
3965.4, \FeI\ $\lambda$ 3966.1, \FeI\ $\lambda$ 3966.6, \FeI\
$\lambda$ 3969.3, \CaIIH\ and \HeI. Solid lines represent the
Gaussian fit; its width is indicated in each plot.}
\label{fig:velocity_hist}
\end{figure}

The left panels of Fig. \ref{fig:velocity_hist} show the
histograms of the LOS velocities in the umbra of the sunspot
obtained from all the spectral lines used in this study. We fitted
a Gaussian to the distributions to estimate the root-mean-square
(rms) velocity value. The histograms are sorted from bottom to top
with increasing rms velocity, with \SiI\ having the smallest and
\HeI\ the largest rms velocity. The \FeI\ lines
at 3966.0 and 3966.6 \AA\ have an identical rms velocity value, as
they are formed at close heights. This is also suggested by their
nearly identical line depth (see Fig. \ref{fig:profile_Ca}). From
these histograms and the variation of the rms velocities, we
obtain the first estimate of the relative formation height of the
spectral lines since we expect a monotonic increase of the rms velocity with height.

For comparison, the right panels of Fig. \ref{fig:velocity_hist}
show the histograms of LOS velocities for the same lines in a
region of the quiet Sun. The \HeI\ histogram is not shown because
the \HeI\ absorption in quiet Sun is too low to retrieve a
velocity. The quiet Sun histograms also show the increase of the rms
velocity with the formation height, although the order of the rms
velocity slightly differs from the one obtained in the sunspot,
since all the \FeI\ lines form in a thin layer. As the oscillatory
power in sunspots is suppressed \citep[see, for
example,][]{Hindman+etal1997}, the rms velocities in the quiet Sun
are larger.
Another possible source of the broadening of the quiet Sun
velocity histogram can come from the granulation velocities not
present inside the umbra. We did not remove the granular component
from the velocity variations in the quiet Sun and it can
contribute to some extent to the overall rms velocity variations
\citep[see, \eg\ ][]{Kostyk+Khomenko2002}.

\begin{figure}
\centering
\includegraphics[width=9cm]{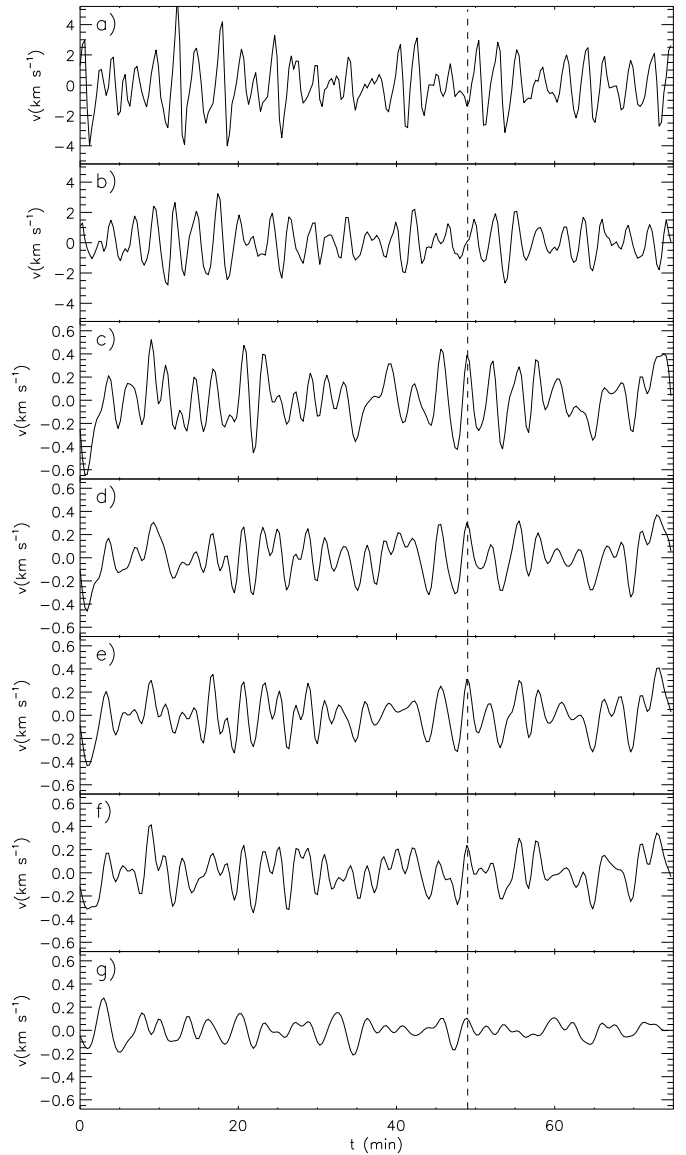}
\caption{Velocity at a fixed position in the umbra for different
spectral lines, sorted by formation height: \HeI\ $\lambda$ 10830
(a), \CaIIH\ 3968.5 (b), \FeI\ $\lambda$ 3969.3 (c), \FeI\
$\lambda$ 3966.6 (d), \FeI\ $\lambda$ 3966.1 (e),\FeI\ $\lambda$
3965.4 (f), \SiI\ $\lambda$ 10827.1 (g). The vertical dashed line
at $t=49$ min marks a prominent velocity peak.}
\label{fig:velocity_x}
\end{figure}

Figure \ref{fig:velocity_x} shows an example of the temporal
evolution of the LOS velocity obtained from the chromospheric
\HeI\ line and the \CaIIH\ line core, the four photospheric \FeI\
lines, and the photospheric \SiI\ line at one location inside the
umbra of the sunspot, indicated in all panels of
Fig.~\ref{fig:velocity_maps} with a vertical dashed line at
$x=28''$. The plots are sorted from bottom to top with increasing
formation height, as retrieved from Fig. \ref{fig:velocity_hist}.
A comparison between the bottom and top panels (which show the
\SiI\ and \HeI\ velocities, respectively) reveals the differences
in period and amplitude of the waves at photospheric and
chromospheric heights. Between these two layers, the rest of the
spectral lines sample different heights of the atmosphere. At
photospheric heights (panels c-g), the pattern of waves is similar
and the signals of the lines with higher formation heights are
slightly delayed (see \eg\ the peak at $t=49$ min marked with
a dashed line). We can see that higher layers have larger
amplitudes (note that the top two panels have a different scale
for the velocity). The temporal evolution of the LOS velocity of
the two chromospheric lines (panels a-b) is almost identical, but
the amplitudes are higher in the case of the \HeI\ line. For
instance from minute 10 to 25, it is clearly seen that the
oscillations measured with the \HeI\ line and \CaIIH\ core have a
saw-tooth profile that indicates the presence of a shock wave
train, with a slow increase of the velocity followed by a sudden
decrease. There is a phase difference of about 20 s between the
\CaIIH\ line core and the \HeI\ line velocities, in the sense that
the oscillatory signal reaches the formation height of the \CaIIH\
line core 20 s before that of the \HeI\ line.

\begin{table}[hbp]
  \begin{center}
  \centering \caption{Rms velocities in ms$^{-1}$ in the sunspot and the
      quiet Sun, and their ratio (3rd column).}
  \label{tb:rms_ratio}
  \smallskip
  \begin{tabular}{cccc}
line             & umbra&  QS & ratio QS/umbra\cr\hline
\SiI\            &   90    &   250     &   2.80    \\
\FeI\ 3965.4     &   170        &  256      &   1.51    \\
\FeI\ 3966.1     &   174        &  275      &   1.58    \\
\FeI\ 3966.6     &   180        &  273      &   1.52    \\
\FeI\ 3969.3     &   201        &  306      &   1.53    \\
\CaIIH\          &   510        &  659      &   1.29    \\
\HeI\            &   726        &  --    &    --  \\
\hline
 \end{tabular}
  \end{center}
 \end{table}

Table \ref{tb:rms_ratio} lists the rms velocities in the quiet Sun
(except \HeI) and in the sunspot for all the spectral lines, together with the ratios between the quiet Sun and the sunspot
velocities.
The relative increase of the rms velocity of the quiet Sun to the
rms velocity of the sunspot decreases with the height in the
atmosphere, from 2.80 at the photospheric height of the formation
of \SiI\ to 1.29 at the chromospheric height of the formation of
\CaIIH\ line core. The rate of the increase of the rms velocity in
the sunspot is faster, and the difference between velocities from lines
formed at higher layers is smaller. The growth of the amplitude of
the oscillations with height is scaled by the pressure scale
height, $H_0=RT/\mu g$, where $T$ is the temperature, $R$ is the
gas constant, $g$ is the gravity and $\mu$ is the mean weight of
the atoms. In the umbra the temperature is lower and $H_0$ is
smaller than in the quiet Sun atmosphere, so the amplitude of
umbral oscillations rises faster.

\subsection{Power Spectra}
\begin{figure}
\centering
\includegraphics[width=9cm]{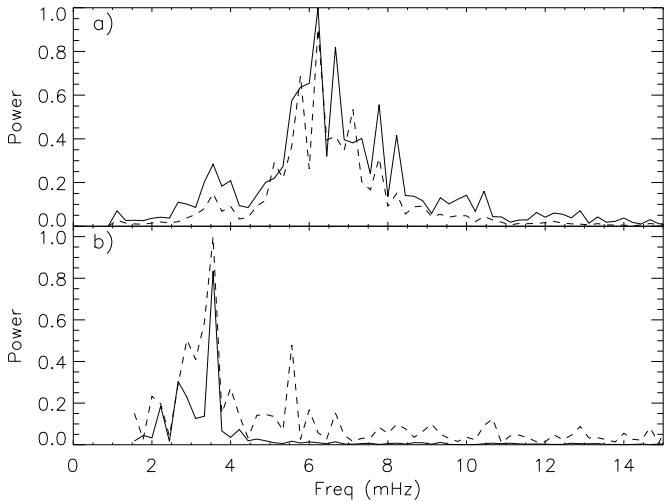}
\caption{Average umbral power spectra of the LOS velocities.
\emph{Top:} chromospheric lines, \HeI\ (\emph{solid line}) and
\CaIIH\ (\emph{dashed line}), both are normalized to the maximum power of the \HeI\ line; \emph{bottom:} photospheric lines,
\SiI\ line (\emph{solid line}) and \FeI\ 3969.3 \AA\ (\emph{dashed
line}), both are normalized to the maximum power of the \FeI\ line.}
 \label{fig:power_spectra}
\end{figure}

Figure \ref{fig:power_spectra} shows the normalized average power
spectra of LOS velocity of the two chromospheric (\HeI\ and
\CaIIH) and two of the photospheric (\SiI\ and \FeI\ $\lambda$
3969.3) lines inside the umbra of the sunspot. We chose this iron
line because it has better signal to noise and its formation
height is distant from the layer where \SiI\ is formed. In the photosphere (bottom panel), the power is concentrated between 2 and 4 mHz, corresponding to the 5 minute band, with a maximum peak at 3.5 mHz. Both spectral lines peak at the same frequency, although the power of the \FeI\ line is slightly higher. The increase of the power at frequencies above 4.5 mHz is more important than the one for frequencies below this value. 

The velocity power spectra of both chromospheric lines (top panel)
have a broad distribution of frequencies, with the largest power
being in the band from 5 to 10 mHz. The chromospheric power
spectrum has a maximum at 6.2 mHz and several secondary peaks
around it \citep[see for comparison][]{Lites1986}. These frequency
peaks correspond to the chromospheric 3 minutes oscillations. At the highest peak of the power spectra, both \CaIIH\ and \HeI\ have almost the same power, but for those frequencies with lower power, the power of the \HeI\ is increased comparing to the \CaIIH. Note
that at heights sampled by our spectral lines we do not find a
continuous transition from the peak at 3.5  mHz to the one at 6.2
mHz, but rather a discontinuous behavior between the photospheric
and chromospheric power spectra. However, the prominent secondary peak around 5.5 mHz in the power spectra of the \FeI\ line is much more obvious than the corresponding in the \SiI\ power spectra, which could indicate some transition towards higher frequencies in the power spectra as the waves propagate upward from the formation height of the \SiI\ line to the \FeI\ lines.

\subsection{Phase spectra}
\label{sect:phase_spectras} A phase diagram gives the phase
difference ($\Delta \phi$) between two signals. In our study, we
use $\Delta \phi$ to measure the time delay between the
oscillatory velocity signals from two spectral lines and assume
that the difference between them is mainly due to the difference
of the formation height of the two lines. In the following, we
will show the phase difference spectra between different
combinations of pairs of spectral lines used in this work. To
obtain the phase spectra, we treated each spatial point separately and
calculated the Fourier-transform of the temporal evolution of the
respective velocities. We derived the phases, and from them the
phase difference of the two signals as a function of the
frequency. There is a 2$\pi$ ambiguity in the computation of the
phase value, so all phase differences have been projected in the
range $\pm \pi$. Then we calculated histograms of the relative
occurrence of a given value of the phase differences at each
frequency taking into account all the corresponding spatial points
\citep[see also][ and references therein]{Krijger+etal2001}. We
obtained the data displayed in
Figs.~\ref{fig:dfase_SiHe}--\ref{fig:dfase_FeSi_quiet}.

In addition to the phase difference spectra, we calculated the
coherence spectra. They provide an estimate of the statistical
validity of the phase and power spectra. Considering $n$ pairs of
signals $x_k(t)$ and $y_k(t)$, whose Fourier transforms are
$\bar{X}_k(\omega)$ and $\bar{Y}_k(\omega)$, respectively, the
coherence is defined as

\begin{equation}
P_{xy}(\omega)=\frac{\frac{1}{n}\Big |\sum_k^n|\bar{X}_k(\omega)||\bar{Y}_k(\omega)|e^{i\Delta\phi_k(\omega) }\Big |}{\frac{1}{n}\sum_k^n\sqrt{|\bar{X}_k(\omega)|^2|\bar{Y}_k(\omega)|^2}}\label{eq:coherence}
\end{equation}

\noindent where
$\Delta\phi_k(\omega)=\phi_{xk}(\omega)-\phi_{yk}(\omega)$.
In our case, the sub-index $k$ covers the spatial position. The
coherence evaluates statistically for every frequency $\omega$ the
relation of the $\Delta\phi_k(\omega)$ for the $n$ $k$-signals. It
takes the value 1 when $\Delta\phi_k(\omega)$ is the same for all
the $k$. If the phase difference of the different $k$ is
arbitrary, the coherence takes very low values. We selected a
confidence limit at 0.7, and for frequencies with coherence above
this value we consider the phase spectra to be reliable.

We also analyzed the increase of the amplitude of the
oscillations. We calculated the amplification spectra as the ratio
between the power at two layers, both of them averaged all over
the umbra:

\begin{equation}
A_{xy}=\frac{\sum_k^n|\bar{Y}_k(\omega)|^2}{\sum_k^n|\bar{X}_k(\omega)|^2}.
\end{equation}

\subsubsection{Theoretical model}
\label{sect:theory}

Following \citet{Centeno+etal2006}, the observations were compared with a
model of linear vertical propagation of slow magneto-acoustic wave in an
isothermal atmosphere that includes radiative losses described by Newton's
cooling law. Assuming that the amplitude of the vertical velocity
changes with height by

\begin{equation}
 V(z)=V_0e^{z/(2H_0)}e^{ik_zz}, \label{eq:amplitud}
\end{equation}

\noindent the dispersion relation for such waves is

\begin{equation}
 k_z^2=\frac{\omega^2-\hat{\omega}_{ac}^2}{\hat{c}^2}, \label{eq:dispersion}
\end{equation}

\noindent where
\begin{equation}
 \hat{\omega}_{ac}=\hat{c}/2H_0,\hat{c}^2=\hat{\gamma}gH_0,\hat{\gamma}=\frac{1-\gamma i\omega\tau_R}{1-i\omega\tau_R}, \label{eq:def}
\end{equation}

\noindent and $\tau_R$ is the radiative cooling time for an
optically thin perturbation \citep{Spiegel1957}:

\begin{equation}
\tau_R=\rho c_v/(16\chi\sigma_RT^3). \label{eq:tau}
\end{equation}

The phase difference between oscillations at two heights is
calculated as the difference of the imaginary part of the argument
of the complex exponential in Eq.~\ref{eq:amplitud}, that is,
$\Delta \phi=k_R\Delta z$, where $\Delta z$ is the geometric
distance between the two heights and $k_R$ is the real part of
$k_z$. The amplification spectrum is given by the ratio between
the amplitude at the two layers and is obtained as
$A=e^{2(1/(2H_0)-k_I)\Delta z}$, with $k_I$ being the imaginary
part of $k_z$.

This model allows us to fit the phase and amplification spectra
with three free parameters: the temperature of the atmosphere,
$T$, the difference in height between two lines, $\Delta z$, and
the typical time scale in which the temperature fluctuations are
damped radiatively, $\tau_R$. These free parameters are manually
tuned to match the effective cut-off frequency and the slope
(including its variations) of the phase difference spectra in the
regime of propagating waves above the cut-off frequency.

\subsubsection{\SiI-\HeI\ phase spectra}

\begin{figure}
\centering
\includegraphics[width=9cm]{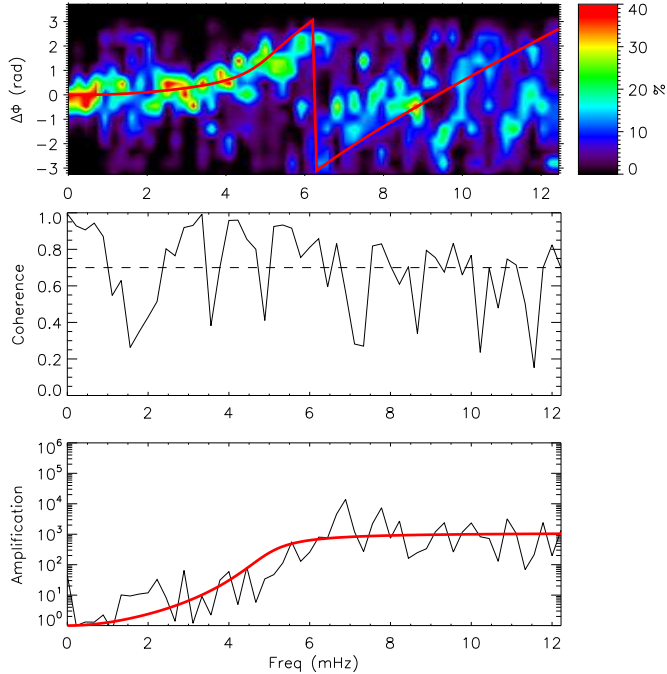}
\caption{\emph{Top}: Phase spectra between the LOS velocities of
the \SiI\ and the \HeI\ lines. The color code shows the relative
occurrence of a given phase shift. The red line represents the
best fit from the theoretical model. \emph{Middle}: Coherence
spectra. The horizontal dashed line at 0.7 marks the confidence
limit. \emph{Bottom}: Amplification spectra. The red line
represents the best fit from the theoretical model.}
\label{fig:dfase_SiHe}
\end{figure}

Figure \ref{fig:dfase_SiHe} shows the phase difference between the
velocity signals measured in the photospheric \SiI\ line and the chromospheric
\HeI\ line. The phase difference is zero for frequencies below 4
mHz. At these frequencies the coherence is high, except in the range between 1 and 2.5 mHz, and the atmosphere oscillates as a whole, \ie, the waves
are stationary. From 4 mHz to 7 mHz, the phase difference increases
linearly with the frequency and the coherence is (more or less)
above the confidence limit. It indicates that waves at these
frequencies propagate from the photospheric layer, where the \SiI\
line forms, to the chromospheric layer, where \HeI\ forms. The
phase difference spectra for higher frequencies are very noisy,
and no meaningful conclusions are possible.

The bottom panel of Fig. \ref{fig:dfase_SiHe} shows the ratio of
chromospheric to photospheric power as a function of frequency. We
fit both the phase difference and the amplification spectra
simultaneously. The parameters of the fit are listed in Table
\ref{tb:parameters}. The solid red line in the phase diagram in
Fig. (\ref{fig:dfase_SiHe}) represents the phase difference
according to the model presented in Sect. \ref{sect:theory},
calculated with the parameters that best fit the observations. The
theoretical amplification spectra matches rather well to the
observational one, with an order of magnitude agreement in the
amplification factor.

\begin{table}[htbp]
  \begin{center}
  \centering \caption{Best-fit parameters of the theoretical model}
  \label{tb:parameters}
  \smallskip
  \begin{tabular}{cccc}
\hline
     Line pair                    & T (K)  & $\Delta z$ (km) & $\tau_R$ (s) \\
\hline
Si-He                    & 4500   &    900          &     45\\
Si-Ca                    & 4000   &    650          &     45\\
Fe 3969.3-He             & 4000   &    450          &     45\\
Fe 3969.3-Si             & 4000   &   -280          &     30 \\
Fe 3969.3-Si (quiet sun) & 4500   &   -200          &     30 \\
Fe 3965.4-Fe 3969.3      & 3500   &    30           &     45 \\
Fe 3966.0-Fe 3969.3      & 3500   &    20           &     45 \\
Fe 3965.4-Fe 3966.0      & 3500   &    10           &     45 \\
Ca-He                    & 6000   &    100          &     15 \\

\hline
 \end{tabular}
  \end{center}
 \end{table}

\subsubsection{\SiI -\CaIIH\ phase spectra}

\begin{figure}
\centering
\includegraphics[width=9cm]{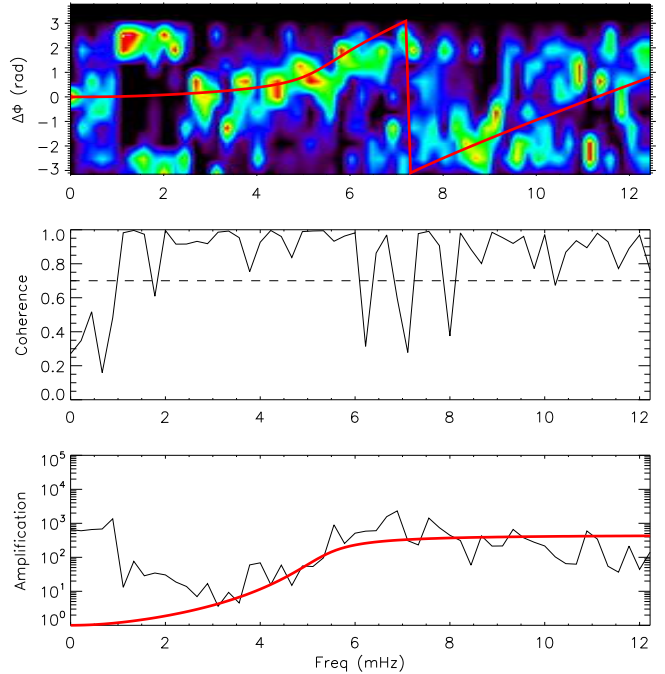}
\caption{Phase, coherence and amplification spectra between \SiI\
and \CaIIH. The format of the figure is the same as Fig.
\ref{fig:dfase_SiHe}.} \label{fig:dfase_SiCa}
\end{figure}

Figure \ref{fig:dfase_SiCa} shows the phase, coherence and
amplification  spectra between the velocity measured with the
\SiI\ line and the core of \CaIIH\ line. Frequencies below 1 mHz have a very low coherence, so these values are not reliable. Surprisingly, frequencies in the range between 1 and 2.5 mHz present an anomalous  behavior, with a phase
difference around 2.5 rad and a high amplification, while their coherence is remarkably high. The long period of waves with these frequencies hinders their analysis, so further observations with longer temporal series are required to study this behavior. The rest of the phase spectra is similar to the one between \SiI\
and \HeI, with zero phase difference between 2.5 and 4 mHz,
indicating stationary waves, and an almost linear increase of the
phase difference between 4 and 7 mHz, corresponding to upwards
propagating waves. In all this frequency range, the coherence is
high. However, in this case the slope of the increase of the phase
difference is smaller than in the previous one, and the
amplification also has lower values. It means that the phase delay
between the velocities at the formation height of these two lines
is smaller than the delay between \SiI\ and \HeI\, and the
amplitude of the waves at the formation height of \HeI\ is larger
than that of waves at the layer where \CaIIH\ forms. Both the
phase and amplification spectra locate the formation height of the
\CaIIH\ core below the \HeI\ line.

The parameters retrieved from the fit of the phase and
amplification spectra to the theoretical model are listed in Table
\ref{tb:parameters}. Comparing the ones retrieved for the line
pair \SiI-\HeI\ with the pair \SiI-\CaIIH, the \CaIIH\ core
formation height in the umbra is around 250 km below the \HeI\
line. The temperature obtained for the phase difference between
\SiI\ and \CaIIH\ is smaller by 500 K; waves traveling from the
formation height of \CaIIH\ core to that of \HeI\ pass presumably
through a region where the temperature is increasing.

\subsubsection{\FeI\ phase spectra}

\begin{figure}
\centering
\includegraphics[width=9cm]{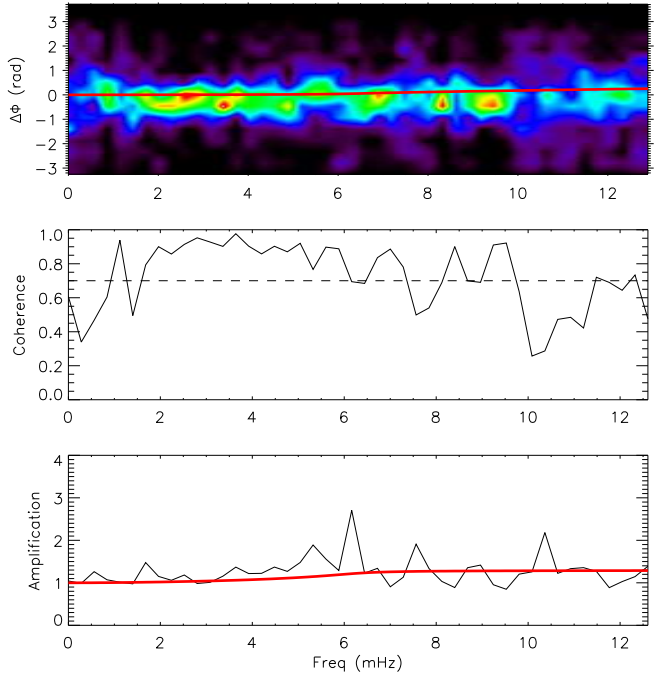}
\caption{Phase, coherence and amplification spectra between \FeI\
$\lambda$ 3966.0 and \FeI\ $\lambda$ 3969.3. The format of the
figure is the same as Fig. \ref{fig:dfase_SiHe}.}
\label{fig:dfase_Fe2Fe5}
\end{figure}

To study the properties of oscillations at photospheric heights,
phase diagrams between pairs of the photospheric lines were
calculated. We can assume that the \FeI\ lines in the \CaIIH\ wing
form at three different heights according to their line depth
(Fig.~3) and to the width of the velocity histograms (Fig.
\ref{fig:velocity_hist}). We take the lines \FeI\ $\lambda$
3965.4, \FeI\ $\lambda$ 3966.0 and \FeI\ $\lambda$ 3969.3 as
representative of these heights, since \FeI\ $\lambda$ 3966.6 and
\FeI\ $\lambda$ 3967.4 seem to form at a similar height as \FeI\
$\lambda$ 3966.0. As an example, Fig. \ref{fig:dfase_Fe2Fe5} shows
the phase spectra, coherence and amplification spectra obtained
between between \FeI\ $\lambda$ 3966.0 and \FeI\ $\lambda$ 3969.3.
The phase spectrum shows that the phase difference is almost zero
for all the frequencies that can be trusted according to the
coherence spectra (from 0 to 10 mHz). We conclude that we can not
retrieve the phase shift between the \FeI\ lines reliably, as the
geometrical difference in their heights of the formation is too
small. However, the amplification spectra reflects some increase
of the amplitude with height. We managed to fit both the phase and
the amplification spectra with the wave model described in the
previous section. These fits yield, indeed, a small geometrical
height difference between the \FeI\ lines. The temperature and the
cooling time obtained from the fit are identical for all pairs of
iron lines (Table \ref{tb:parameters}).

\subsubsection{\FeI -\SiI\ phase spectra}

\begin{figure}
\centering
\includegraphics[width=9cm]{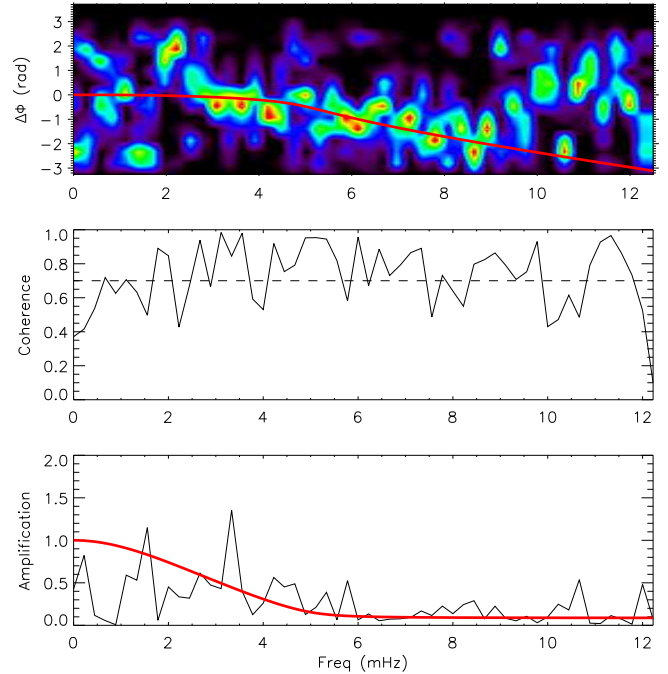}
\caption{Phase, coherence and amplification spectra between \FeI\
3969.3 and \SiI. The format of the figure is the same as Fig.
\ref{fig:dfase_SiHe}.} \label{fig:dfase_FeSi}
\end{figure}

The phase difference spectrum between the \FeI\ $\lambda$ 3969.3
line and the \SiI\ $\lambda$ 10827 line is shown in the top panel
of Fig. \ref{fig:dfase_FeSi}. For frequencies below 2 mHz, the
phase spectrum is very noisy and has no coherence (see middle
panel), indicating that waves at the heights of formation of these
two lines are not related. At frequencies in the band of $2.5-4$
mHz, some oscillatory power is present, the coherence is high and
the phase difference is close to zero, so these waves are
evanescent. The phase difference spectrum for high frequencies in
the range from 4 to 9 mHz shows a decreasing tendency, indicating
that the waves reach the formation height of the \SiI\ line before
the \FeI\ line. This phase spectrum was also fitted with the wave
propagation model. The solid red line in top and bottom panels of
Fig. \ref{fig:dfase_FeSi} shows the result of this fit. The
amplification spectrum is not reliable at frequencies below 3 mHz
due to the low coherence, but above this value the agreement of
the theoretical and observational spectra is good, indicating that
the amplitude of the oscillations measured in \SiI\ is about twice
lower than the amplitude measured in the \FeI\ 3969.3 line.

As a summary of the propagation at photospheric heights, we conclude
  that all observed photospheric spectral lines (five \FeI\ lines and the
\SiI\ line) fluctuate with a dominant period of 5 minutes. While
all the \FeI\ lines have similar velocity amplitudes, with a
maximum peak-to-peak amplitude of around 800 m s$^{-1}$, the
velocity amplitude measured in \SiI\ is clearly smaller (maximum
peak-to-peak amplitude around 400 m s$^{-1}$). According to the
phase diagram between the \FeI\ lines (Fig.
\ref{fig:dfase_Fe2Fe5}) and the amplification and phase diagram
between a \FeI\ line and \SiI\ (Fig. \ref{fig:dfase_FeSi}), we
conclude that waves with frequencies above 4 mHz propagate upwards
at photospheric heights. They first reach the height where \SiI\
is formed. As they propagate upward, their amplitudes increase due
to the density fall-off. Then the waves reach the formation height
of all the \FeI\ spectral lines, that all come from a thin layer
and for this reason all show similar amplitudes.

\subsubsection{\FeI -\HeI\ phase spectra}

\begin{figure}
\centering
\includegraphics[width=9cm]{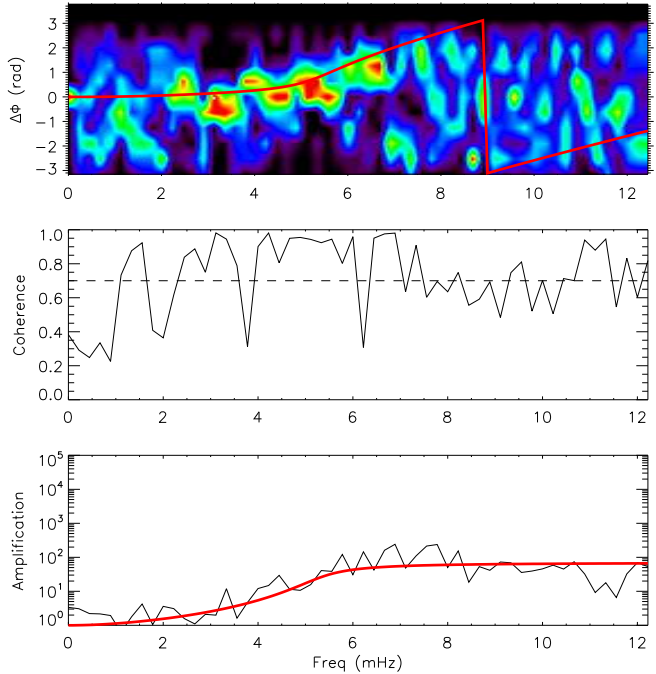}
\caption{Phase, coherence and amplification spectra between \FeI\
$\lambda$ 3969.3 and \HeI. The format of the figure is the same as
Fig. \ref{fig:dfase_SiHe}.} \label{fig:dfase_FeHe}
\end{figure}

According to the results extracted from the velocity statistics
and the phase spectra presented before, the \FeI\ lines form at
some height between the formation height of the \SiI\ line and the
chromospheric \HeI\ line and the \CaIIH\ core. This means that
they give information about a high photospheric layer, located at
an intermediate height in the propagation of the waves from the
photosphere to the chromosphere. Figure \ref{fig:dfase_FeHe} shows
the analysis of the phase differences and amplification between
the velocity signal measured at the formation height of the \FeI\
$\lambda$ 3969.3 line and the \HeI\ line. The phase spectrum is
similar to the other two between photospheric and chromospheric
lines (\SiI -\HeI\ and \SiI-\CaIIH). For frequencies between 2 and
4 mHz, the phase difference is about zero, and it increases with
the frequency between 4 and 7 mHz, but the slope is smaller than
for \SiI-\HeI. In the range between 2 and 7 mHz, the coherence is
significant. The amplification spectrum also shows a lower
amplification, compared to the \SiI-\HeI\ case. The temperature,
height difference and radiative damping time retrieved from the
fit are listed in Table \ref{tb:parameters}.

\subsubsection{\CaIIH\ -\HeI\ phase spectra}

\begin{figure}
\centering
\includegraphics[width=9cm]{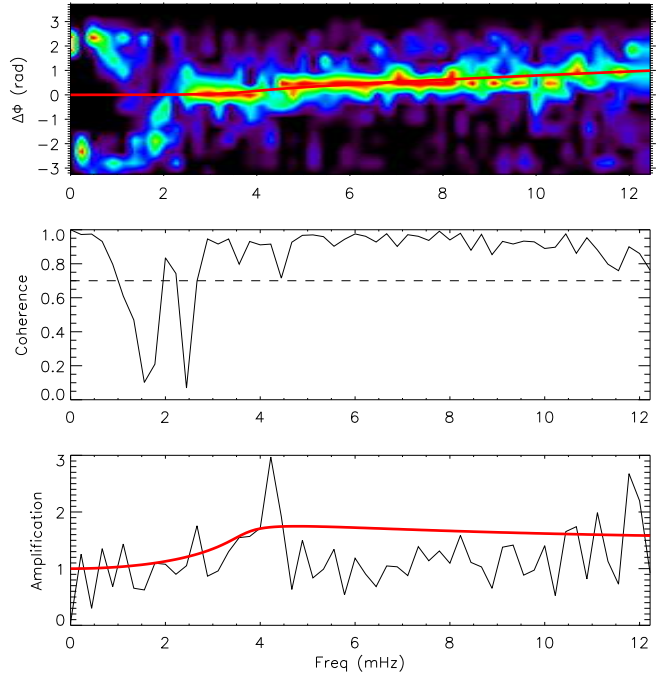}
\caption{Phase, coherence and amplification spectra between
\CaIIH\ and \HeI. The format of the figure is the same as Fig.
\ref{fig:dfase_SiHe}. } \label{fig:dfase_CaHe}
\end{figure}

At chromospheric heights, we have the Doppler velocities obtained
from the cores of the \CaIIH\ line and the \HeI\ $\lambda$ 10830
line. The top panel of Fig. \ref{fig:dfase_CaHe} shows the phase
difference diagram between them. At frequencies between 1-2 mHz,
the coherence (middle  panel of Fig. \ref{fig:dfase_CaHe}) is low
and the phase spectrum is noisy. The coherence spectrum shows that
the phase spectrum is reliable between 2 and 12 mHz. In the
frequency band between 2 and 4 mHz, the phase difference is about
0, indicating that there is no propagation and the waves are
evanescent. From $\nu$=4 mHz to $\nu$=11 mHz, the phase difference
increases, starting from $\Delta\phi=0$ and showing a small
positive slope. It means that as waves propagate upwards, they
reach the \CaIIH\ core formation height just before the \HeI\ one.
The same conclusion was obtained previously from comparing the
temporal variations of the \CaIIH\ core and \HeI\ velocities
presented in Fig.~\ref{fig:velocity_x}(a--b), where the oscillatory
signal of \HeI\ is delayed by 20 s with respect to the \CaIIH\ one.

The differences in the amplitudes between \CaIIH\ core and \HeI\
velocities are in line with their phase spectra (top panel in Fig.
\ref{fig:dfase_CaHe}), since the amplitude of the \HeI\ velocity
is bigger than the \CaIIH. A more detailed inspection of the
amplification spectra between both chromospheric lines (bottom
panel of Fig. \ref{fig:dfase_CaHe}) reveals that the oscillatory
signal is amplified between 2 and 4 mHz, but there is no
amplification for frequencies in the range from 4.5 to 10 mHz  (the ratio between the
amplitudes is around unity).
Waves in  the 2--4 mHz frequency range are evanescent, and the
increase of their amplitude from the photosphere
to the chromosphere is not so high, keeping them in a linear
regime. At the high layers, their amplitude still increases due to
the drop of the density with height. On the other hand, waves with
frequencies between 4  and 10 mHz propagate upwards to the chromosphere and
develop into shocks (top panels of Fig. \ref{fig:velocity_x}).
In this non-linear regime,  their amplitudes do not increase with
height, explaining the observational amplification spectrum
around unity for frequencies in the range 4-10 mHz. The model of linear wave
propagation in an isothermal stratified atmosphere predicts a
higher amplification. In the case of waves with frequencies above 10 mHz, they do propagate upwards, but their amplitude at the photosphere is so low that they do not reach a non-linear regime and their amplitude still increases with height at chromospheric layers. All in all, the model is not suitable for the
description of waves at heights of \CaIIH\ core and \HeI\
formation where the non-linearities start to become important.

It is expected that the propagation of non-linear waves happens at
a higher speed compared to the linear case. This would decrease the
phase difference between two layers. Therefore, we can expect
that the height difference between the formation layers of
\CaIIH\ and \HeI\ lines presented in Table \ref{tb:parameters} is
underestimated by our linear model; the value of about 100 km is
the lower limit of this difference. One of the issues that arises
from this fact is the evaluation of the height range where the
linear model of wave propagation is valid. It is clear that it
fails between the formation height of the \CaIIH\ core and the
\HeI\ line due to the non-linearities. However, in previous
sections we have applied successfully this model to the
propagation between the photospheric \SiI\ line and the
chromospheric \CaIIH\ core and \HeI\ line, and between one of the
\FeI\ lines and the \HeI\ line. It means that in most of their way
from the photosphere to the chromosphere, the linear regime is a
good approximation for these waves. At some layer between the
formation height of the \FeI\ 3969.3 and the \HeI\ lines the wave
propagation departs from the linear regime
\citep[cf.][]{carlsson+stein1997}.

\subsubsection{\FeI-\SiI\ phase spectra in quiet Sun}

\begin{figure}
\centering
\includegraphics[width=9cm]{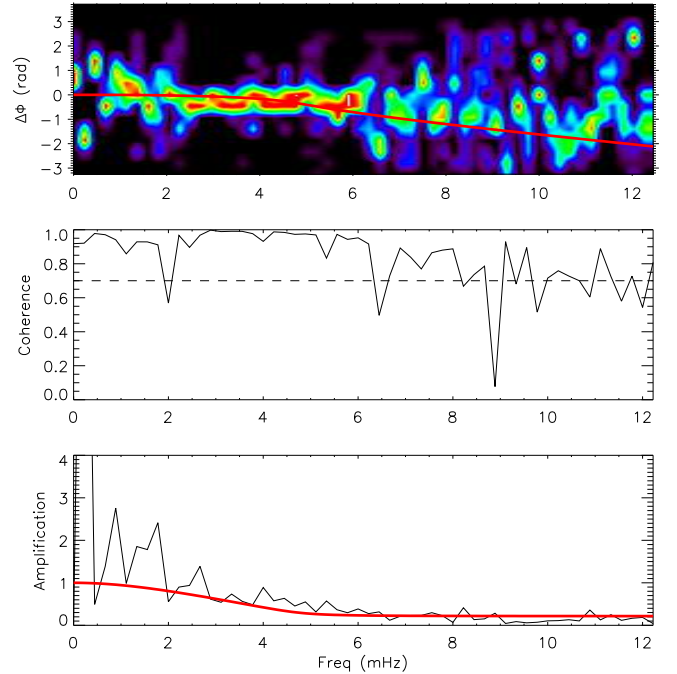}
\caption{Phase, coherence and amplification spectra between \FeI\
$\lambda$ 3969.3 and \SiI\ in a quiet Sun region. The format of
the figure is the same as Fig. \ref{fig:dfase_SiHe}.}
\label{fig:dfase_FeSi_quiet}
\end{figure}

We calculated the phase difference spectra for a region of quiet
Sun around the sunspot as well. In this region, the Stokes
parameters $QUV$ are below the level of noise, so we suppose there
is no significant magnetic field. Figure
\ref{fig:dfase_FeSi_quiet} (top) shows the phase spectrum between
the photospheric \FeI\ $\lambda$ 3969.3 and \SiI\ $\lambda$ 10827
lines in the quiet Sun. The coherence spectrum shows high values
for frequencies below 9 mHz. For frequencies below 2 mHz, the
phase diagram is very noisy due to the low oscillatory power of
the two velocity signals, and we do not find any clear relation
between them. In the spectral range from 2 mHz to 4 mHz, the phase
difference takes a constant value around zero, while for
frequencies above 4 mHz, $\Delta \phi$ decreases with increasing
frequency. However, the low power at the photosphere at
frequencies between 6 and 8 mHz (Fig. \ref{fig:power_spectra})
results in a noisy phase spectrum in this band.

Comparing the phase difference diagrams between the same two lines
inside the umbra of the sunspot (top panel in Fig.
\ref{fig:dfase_FeSi}) and in the quiet Sun, we find that in quiet
Sun the negative slope of the phase difference is less steep. We
also fitted the quiet Sun phase difference and amplification
spectra with our wave propagation model. Inside the umbra of the
sunspot, the magnetic field imposes wave propagation along field
lines, and thus the hypothesis of vertical propagation is
justified. However, in the quiet Sun waves can propagate in
different directions and this approximation may not be true.
Still, we were able to find a fit matching reasonably the phase
spectrum. We retrieve a lower value of formation height
differences and a higher temperature in the quiet Sun compared to
the umbra (see Table \ref{tb:parameters}).

The bottom panel of Fig. \ref{fig:dfase_FeSi_quiet} shows the
amplification spectrum for this case. For frequencies in the band
of 2-7 mHz, the theoretical amplification matches the
observational one.

\subsection{Comparison of the parameters of the fit with a model of sunspot}
\label{sect:sunspot_model}

From the fit to the phase and amplitude spectra of all the pairs
of lines we obtain the temperature, the difference between the
formation heights of two lines, and the cooling time that best
match the observational data. Obviously, our model is simplified
and has several important limitations. It only describes linear
wave propagation in an isothermal stratified atmosphere, not
taking into account a realistic temperature stratification of the
sunspot atmosphere. In a real sunspot, the temperature varies with height, so the temperature that we obtain from the fit
represents a mean value between two heights of formation. In the
case of spectral lines formed at a similar height (\ie, the \FeI\
lines), the temperature that we find should be close to the real
temperature in the layer. However, when the formation height
difference ($\Delta z$) obtained from the fit is larger (\ie,
\SiI-\HeI), we can not assign the temperature to a certain height.
The presence of shocks in the temporal evolution of the \HeI\ and
\CaIIH\ velocities (Fig. \ref{fig:velocity_x}), together with the
difficulties of the model to fit the amplification spectra between
\CaIIH\ and \HeI\ (Fig. \ref{fig:dfase_CaHe}), show that the
propagation at these heights is non-linear. Thus our determination
of the height difference between these two layers must be somewhat
affected by the deviations from the linear regime.

\begin{figure}
\centering
\includegraphics[width=9cm]{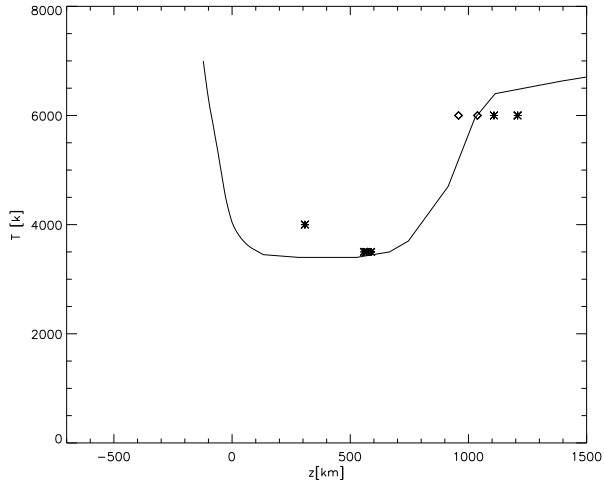}
\caption{Temperature stratification in the sunspot model of
\citet{Maltby+etal1986}. Asterisks represent our best-fit values to the
phase difference and amplification spectra of the line pairs.
Diamonds  mark an alternative estimate of the formation height of
the \CaIIH\ core and the \HeI\ line.} \label{fig:strat_T}
\end{figure}

Despite these limitations, we plotted the deduced temperature
values over the temperature stratification of the sunspot model of
\citet{Maltby+etal1986} (Fig. \ref{fig:strat_T}). The symbols in this
figure mark the values retrieved from the fit to our observations.
From the fit, we obtain the relative difference between the
formation heights of spectral lines, not the absolute values. As
reference point, we set the height of the \SiI\ line to z=308 km, as given by \citet{bard+carlsson2008} for the sunspot atmosphere model of \citet{Maltby+etal1986}. The
formation heights of the other lines then follow from their
relative distance to the \SiI\ line (Table \ref{tb:parameters}).
Table \ref{tb:formation} lists the resulting height values and the
corresponding temperature.

In the case of the \FeI\ lines, we obtained
from the low phase difference (top panel of Fig.
\ref{fig:dfase_Fe2Fe5}) that the distances between them are very
small, but the amplification spectra (bottom panel of Fig.
\ref{fig:dfase_Fe2Fe5}) or the histograms of velocities (Fig.
\ref{fig:velocity_hist}) show a certain increase of the amplitude
with height. This amplification in spite of the small geometrical
distance indicates a low pressure scale height $H_0$ and
consequently a low temperature $T$. The height of the \FeI\ line
close to the temperature minimum agrees with the results from the
fit of the model to the phase and amplification spectra.

In the case of the chromospheric signals (\CaIIH\ core and \HeI\
line), we assign to both formation heights the temperature
retrieved from the fit of the phase spectra between \CaIIH\ and
\HeI\ velocities. According to the geometrical differences of
Table \ref{tb:parameters}, there is an uncertainty in
the formation height of these two lines. On the one hand, we have
set them from their $\Delta z$ with respect to the \SiI\ line,
locating the formation height of the \CaIIH\ core at 958 km and
the formation height of the \HeI\ line at 1208 km. On the other
hand, we have obtained an alternative height for the \CaIIH\ core
by considering its $\Delta z$ to the \HeI\ line, and subtracting
it from the previous estimate of the \HeI\ formation height. In
the same way, we have located the formation height of the \HeI\
line taking into account the geometrical differences between the
pairs \SiI-\FeI\ $\lambda$ 3969.3 and \FeI\ $\lambda$ 3969.3-\HeI.
Thus, we retrieved a range of heights for both chromospheric
lines, as it is shown in Table \ref{tb:formation}. The height
ranges we find are comparable to previous works
\citep[e.g.,][]{lites+etal1993, carlsson+stein1997, centeno+etal2009, beck+etal2009}, even if some
of these articles deal with the quiet Sun solely.

\begin{table}[htbp]
  \begin{center}
  \centering \caption{Formation heights and temperature for the spectral lines}
  \label{tb:formation}
  \smallskip
  \begin{tabular}{ccccc}
\hline
Spectral line    & z [km]  & T[k]    \\
\hline
\SiI\            & 308      & 4000    \\
\FeI\ 3965.4     & 558      & 3500    \\
\FeI\ 3966.0     & 568      & 3500    \\
\FeI\ 3969.3     & 588      & 3500    \\
\CaIIH\          & 958-1108  & 6000    \\
\HeI\            & 1038-1208 & 6000    \\
\hline
 \end{tabular}
  \end{center}
 \end{table}

\section{Discussion and conclusions}
\label{sect_disc}

We have presented an analysis of the LOS velocities obtained from
at set of spectropolarimetric data in the near-IR spectral region
around 10830 \AA\ and the optical region around 3969 \AA\ in a sunspot
atmosphere and its vicinity. From these two spectral ranges, we
retrieve on the one hand the Doppler velocities of the
photospheric \SiI\ $\lambda$ 10827 and the chromospheric \HeI\
$\lambda$ 10830 line. On the other hand, we also sample several
layers between these two heights, using the Doppler shifts of the
chromospheric \CaIIH\ line core and the photospheric \FeI\ lines
from the wings of the \CaIIH\ line.

The histograms of LOS velocities show that the width of the
velocity distribution increases with height, both in the sunspot
and the quiet Sun atmosphere. Quiet Sun rms velocities are larger
than those in the sunspot due to the higher power of oscillations
in the quiet Sun. The growth of the amplitude of the oscillations
with height is scaled by the pressure scale height $H_0$. It is
smaller in the umbra than in the quiet Sun atmosphere, so the
amplitude of umbral oscillations rises faster. This yields that
the ratio of the quiet Sun and umbral rms velocities of the same
lines decreases with the formation height of the lines.

The phase difference spectra of LOS velocities between several
pairs of lines show upward propagating waves for frequencies
higher than 4 mHz. The power at lower frequencies does not
propagate up, since waves with these frequencies are evanescent.
The slope of the phase spectra, together with the histograms of
LOS velocity, allows us to sort all the spectral lines used in
this work by their formation height.

Phase and amplification spectra were fitted to a model of linear
wave propagation in a stratified atmosphere with radiative losses
following the Newton cooling law. The model works reasonably well
at layers below the formation height of the \CaIIH\ line core,
where waves propagate in a linear regime, while it fails in the
fit of the amplification spectrum  between the \HeI\ velocity and
the \CaIIH\ line core velocity, due to the importance of
nonlinearities at these chromospheric heights.

From the fit to the model, we retrieved the temperature, the
difference in geometrical height between the formation heights of
both spectral lines, and the radiative relaxation time. Setting
the height of the lowermost forming line (\SiI) to agree with \citet{bard+carlsson2008}, the formation height of all the lines in sunspots was
inferred. The \FeI\ lines from the wings of the \CaIIH\ line are formed about 250 km above the photospheric \SiI\ line. The relative position of the lines is well determined by the observations, since the rms velocities (Fig. \ref{fig:velocity_hist}), the power spectra (Fig. \ref{fig:power_spectra}), and the phase and amplification spectra (Fig. \ref{fig:dfase_FeSi}) all indicate that the \FeI\ lines are formed in the upper photosphere above the \SiI\ line. The temperatures obtained for the spectral lines then
show a good agreement with the temperature stratification of the  \citet{Maltby+etal1986} sunspot model, and the formation heights are coherent with previous estimates.  It must, however, be taken into account that the estimate of the formation height from response functions in \citet{bard+carlsson2008} was performed for a static atmosphere and has an strong dependence on the atmosphere model employed.

Most of the power of the photospheric lines is concentrated in a
prominent peak at 3.5 mHz, in the 5 minute band. From a comparison
between the power spectra of the \SiI\ and one of the \FeI\ lines,
it is interesting to note that the power peak is exactly at the
same position, although the iron line forms at around 200 km above
the silicon one. The same behavior was found for the chromospheric
\HeI\ line and the \CaIIH\ line core, which peak at around 6 mHz,
corresponding to the 3 minute band. The maximum of the power
spectra is not shifted gradually from 3.5 mHz at the photosphere
to higher frequencies at larger heights, but the photospheric
and chromospheric group of lines show a discontinuous behavior.
Waves at frequencies above the cut-off increase their amplitude
with height faster than evanescent waves below the cut-off,
resulting in larger power of 3-minute waves at chromospheric
heights.
Yet, as follows from Fig.~\ref{fig:power_spectra}b, the LOS
velocity power at frequencies above 4 mHz measured from  of the
\FeI\ line is higher than the one from the \SiI\ line; it means
that already in the upper photosphere the high-frequency power
becomes important.
This finding suggests that high-frequency waves prominent in the
chromosphere have to be generated in the photosphere or below and
their dominance at the chromospheric height is the result of their
large amplitude increase. These results are consistent with those
obtained from observations by \citet{Centeno+etal2006} using two
\HeI\ and \SiI\ lines, and from numerical simulations by
\citet{Felipe+etal2010a}.

From the compatibility between our observations and a simple
wave model, we conclude that we observe a continuous field-aligned
propagation of slow magneto-acoustic waves in the upper atmosphere
of sunspot. These waves first reach the formation height of \SiI,
then the formation height of the \FeI\ lines from the \CaIIH\ line
wing located in the upper photosphere, then the formation height
of the \CaIIH\ line core and finally that of the \HeI\ line. The
propagation becomes non-linear at heights between the formation of
\FeI\ lines and \CaIIH\ line core.

\acknowledgements  This research has been funded by the Spanish
MICINN through projects AYA2007-63881 and AYA2007-66502.

\aareferences

\end{document}